\documentclass[journal=nalefd,manuscript=letter]{achemso}

\usepackage{siunitx}
\usepackage[version=3]{mhchem} 

\author{Michael Foltýn}
\affiliation[CEITEC]
{Brno University of Technology, Central European Institute of Technology, Purkyňova 123, Brno,
612 00, Czech Republic}

\author{Michal Kvapil}
\affiliation[CEITEC]
{Brno University of Technology, Central European Institute of Technology, Purkyňova 123, Brno,
612 00, Czech Republic}
\alsoaffiliation[UFI]
{Brno University of Technology, Faculty of Mechanical Engineering, Institute of Physical Engineering, Technická 2, Brno, 616 69, Czech Republic}

\author{Tomáš Šikola}
\affiliation[CEITEC]
{Brno University of Technology, Central European Institute of Technology, Purkyňova 123, Brno,
612 00, Czech Republic}
\alsoaffiliation[UFI]
{Brno University of Technology, Faculty of Mechanical Engineering, Institute of Physical Engineering, Technická 2, Brno, 616 69, Czech Republic}

\author{Michal Horák}
\affiliation[CEITEC]
{Brno University of Technology, Central European Institute of Technology, Purkyňova 123, Brno,
612 00, Czech Republic}
\email{michal.horak2@ceitec.vutbr.cz}

\title[] {Plasmonic Properties of Individual\\Bismuth Nanoparticles}

\begin{document}

\begin{abstract}
  
Bismuth nanoparticles are being investigated due to their reported photothermal and photocatalytic properties. In this study, we synthesised spherical bismuth nanoparticles (50–600 nm) and investigated their structural and optical properties at the single particle level using analytical transmission electron microscopy. Our experimental results, supported by numerical simulations, demonstrate that bismuth nanoparticles support localised surface plasmon resonances, which can be tuned from the near-infrared to the ultraviolet spectral region by changing the nanoparticle size. Furthermore, plasmonic resonances demonstrate stability across the entire spectral bandwidth, enhancing the attractiveness of bismuth nanoparticles for applications over a wide spectral range. Bismuth's lower cost, biocompatibility, and oxidation resistance make it a suitable candidate for utilisation, particularly in industrial and large-scale plasmonic applications.
  
\end{abstract}

\vspace{3cm}

The biocompatibility, high atomic number, and accessible functionalisation of bismuth nanoparticles make them a compelling candidate as a contrast enhancing agent in medical X-ray imaging and computational tomography techniques \cite{10.1016/j.jddst.2021.102895, 10.1016/j.jpcs.2018.03.034}. Bismuth nanoparticles also exhibit quantum confinement effects \cite{10.1063/1.2192624}, high Seebeck coefficients \cite{10.1002/anie.201005023}, and thermally driven semimetal to semiconductor transitions \cite{10.1088/0957-4484/21/40/405701}. Furthermore, their recently reported photocatalytic and photothermal properties also make them attractive for use in photothermal cancer therapies \cite{10.1021/acsanm.7b00255, 10.1039/D0CS00031K}, environmental remediation \cite{10.1039/D3EN00983A, 10.1021/jacs.3c04727, 10.1039/C4CC02724H}, and energy storage \cite{10.1016/j.ensm.2023.03.023}. Theoretical studies have also predicted the ability of bismuth to support collective oscillations of free electrons, known as localised surface plasmon resonances (LSPR) \cite{10.1038/nmat2630, 10.1039/C3CP43856B}. These resonances enhance the local electromagnetic field in the vicinity of the nanoparticle, offering rich applications in biosensing \cite{10.1021/acssensors.8b00315, 10.1515/nanoph-2023-0317}, metasurfaces \cite{10.1002/adom.202302130}, and medicine \cite{10.1021/acssensors.4c03562}. The potential combination of plasmonic applications and the extraordinary properties of bismuth has fuelled research on nanostructured bismuth thin films \cite{10.1016/j.photonics.2022.101058, 10.1021/jp3065882, 10.1002/adom.202302130, 10.48550/arXiv.2504.00671, 10.1039/C7CP04359G}.

However, studies on the plasmonic response of chemically synthesised bismuth nanoparticles are limited only to investigations of plasmonic performance using far-field optical spectroscopy \cite{10.1021/jp046423v, 10.1021/jp4065505}. The primary constraint of the method is that it quantifies the response of a large volume of solution containing nanoparticles of various sizes. Therefore, the recorded spectrum comprises contributions from nanoparticles of all sizes, resulting in an overlap of individual plasmonic resonances \cite{10.1021/acs.inorgchem.1c02621}. Ultimately, the dependence of the plasmon energy on the size of the nanoparticles cannot be determined and the plasmonic performance cannot be quantified \cite{10.1155/2013/313081, 10.1021/ac502053s}. In order to obtain this information, an alternative analytical method with sufficient spatial and spectral resolution is required to measure the plasmonic resonances of individual nanoparticles. Electron energy loss spectroscopy (EELS) is an analytical technique frequently used conducted in a scanning transmission electron microscope (STEM) \cite{10.1103/RevModPhys.82.209}. Notwithstanding the advantages inherent in this method, an analysis of the synthesis approach of the nanoparticles is required. The majority of wet nanoparticle syntheses rely on surfactants to prevent the aggregation and oxidation of synthesised nanoparticles. However, the surfactant molecules that are adsorbed onto the analysed nanoparticles introduce an undesirable signal arising from their own excitations and alter the local dielectric environment \cite{10.1002/adom.202101221, 10.3390/polym12071434}. These detrimental effects have the potential to hinder EELS analysis or even prevent the extraction of signals emanating from plasmon resonances \cite{10.1063/1.4965862}. It is imperative that a surfactant-free process is employed during synthesis in order to successfully analyse the plasmonic response of individual bismuth nanoparticles by STEM EELS. In this study, we present the STEM EELS analysis of the optical response of individual monocrystalline bismuth spherical nanoparticles synthesised by a surfactant-free polyol process and demonstrate that the dipole LSPR mode can be tuned from near-infrared (NIR) to ultraviolet (UV) spectral region as a function of the nanoparticle diameter.

\begin{figure}[t]
    \centering
    \includegraphics[width=1\linewidth]{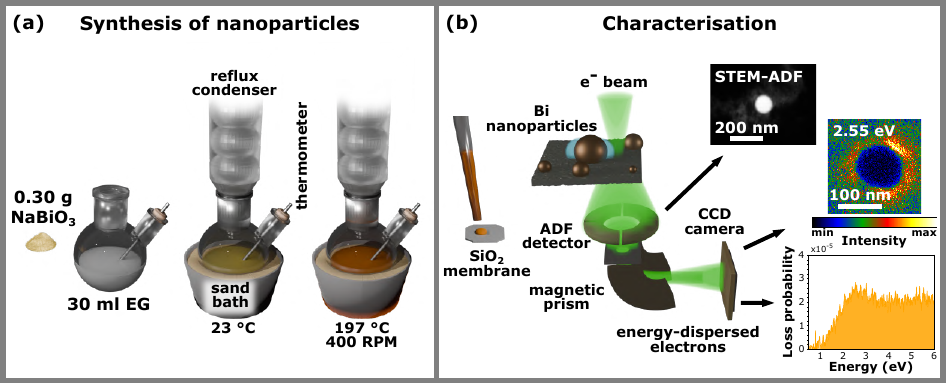}
    \caption{Schematic workflow of the synthesis and characterization of bismuth monocrystalline nanoparticles: (a) The nanoparticles were prepared by reducing sodium bismuthate in ethylene glycol at temperatures near its boiling point. (b) The morphology of the synthesized nanoparticles was obtained from STEM ADF micrographs and their plasmon resonances were examined using STEM EELS.}
    \label{Fig1}
\end{figure}

The schematic workflow for the synthesis and characterisation of bismuth nanoparticles is shown in Figure~\ref{Fig1}. Bismuth nanoparticles were synthesised using a modified polyol process based on the synthesis approach reported in Ref. \cite{10.1021/jp063474e}. The main steps of the reduction of sodium bismuthate in ethylene glycol at temperatures close to its boiling point are depicted in Figure~\ref{Fig1}a. For further details, see Methods. Next, we diluted the solution with methanol and dropped it onto commercially available \ce{SiO2} and carbon membranes for subsequent characterisation by analytical transmission electron microscopy. Figure~\ref{Fig1}b shows the setup used for the STEM EELS analysis of synthesised nanoparticles. It includes a STEM annular dark field (ADF) image of a \SI{107}{\nano\meter} nanoparticle, a background-subtracted electron energy loss spectrum integrated on the left edge of the nanoparticle, with a peak at \SI{2.55}{\electronvolt} corresponding to the dipole LSPR mode, and a loss probability map of electrons with energy loss of \SI{2.55}{\electronvolt}, the energy of the dipole mode.

\begin{figure}[ht!]
    \centering
    \includegraphics[width=1\linewidth]{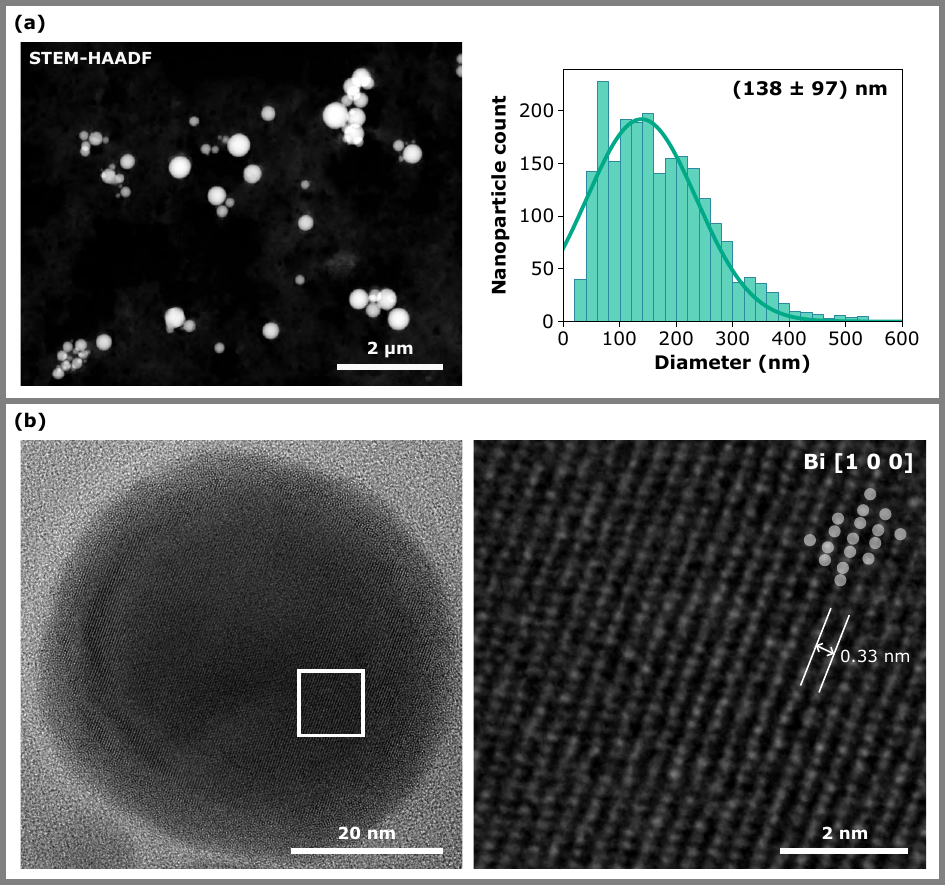}
    \caption{Structure and morphology of synthesised nanoparticles: (a) STEM HAADF micrograph of synthesised bismuth nanoparticles showing their typical shape and their size distributions. The nanoparticle diameter histogram diameter indicates large dispersion in size. (b) High-resolution TEM of a \SI{54}{\nano\meter} nanoparticle showing that the nanoparticle is  monocrystalline in $[1\;0\;0]$ orientation. The white frame indicates the location of the zoomed area.}
    \label{Fig2}
\end{figure}

First, bismuth nanoparticles were inspected by analytical transmission electron microscopy to characterise their size, morphology, crystallinity, and chemical composition. Figure~\ref{Fig2}a shows a typical STEM high-angle annular dark-field (HAADF) micrograph of synthesised nanoparticles. The nanoparticles are generally spherical. However, smaller amounts of nanoparticles of different shapes, including nanowires, hexagonal nanoparticles, truncated nanotriangles, and nanosquares, were obtained, too (see Figure~S1). The absence of surfactant molecules results in a rather wide-diameter distribution of the synthesised spherical nanoparticles. Based on the Gaussian fit of the diameter histogram, shown in Figure~\ref{Fig2}a, the average diameter of the nanoparticles is $(137 \pm 97)$, nm. We found that as the synthesis time is further increased (the time for which the temperature is kept at \SI{197}{\celsius}), the average diameter of the nanoparticle also increases (see Figure~S2). This increase can be explained by the slow gradual reduction of unconsumed residual sodium bismuthate in the solution over time. In addition, the nanoparticles were sometimes embedded in layers of chemical residues (see Figure~S3). Energy-dispersive X-ray spectroscopy (EDX) indicates that these chemical residues are likely unconsumed sodium bismuthate (see Figure~S4). However, scanning the residue with the focused electron beam successfully removed them, likely through reduction by the electron beam. We note that the reduction of sodium bismuthate using an electron beam was reported in Ref. \cite{10.1088/0957-4484/18/33/335604}.

Despite continuous growth during synthesis, the nanoparticles are monocrystalline. Figure~\ref{Fig2}b shows the high-resolution TEM micrograph of a \SI{54}{\nano\meter} nanoparticle that demonstrates its monocrystallinity. Individual atomic columns are resolved. The image was further evaluated and compared with the crystalographic model of rhombohedral bismuth introduced in Ref. \cite{10.1039/C9CC02820J} using CrystBox \cite{10.1107/S1600576717006793}. As a result, the bismuth nanoparticle on the high-resolution TEM micrograph is in the $[1\;0\;0]$ orientation with the distance of \SI{0.33}{\nano\meter} between the atomic planes. In addition, there is no deep oxidation present at the nanoparticle surface despite the absence of surfactants in the synthesis. The surface shell exhibits the properties of a native oxide layer that is just a few nanometres thick, which is further confirmed by the EDX shown in Figure~S4. This is likely due to ethyleneglycol and its low oxidation potential at elevated temperatures \cite{10.1039/C3DT52242C}.

\begin{figure}[t]
    \centering
    \includegraphics[width=1\linewidth]{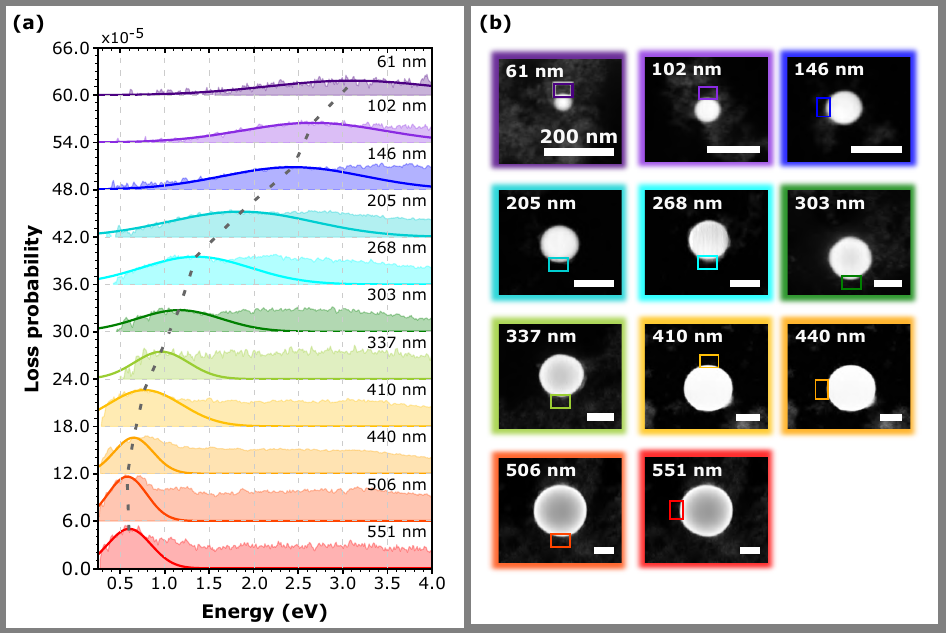}
    \caption{EELS analysis of spherical nanoparticles with diameter ranging from \SI{61}{\nano\meter} to \SI{551}{\nano\meter}: (a) Measured EEL spectra with the peaks corresponding to the dipole LSPR mode fitted with Gaussian. (b) STEM HAADF micrographs of the nanoparticles with rectangles marking the areas from which the EELS signal was collected.}
    \label{Fig3}
\end{figure}

Second, we have focused on the plasmonic properties of bismuth nanoparticles that are measured by STEM EELS. We have investigated a set of spherical nanoparticles with diameters ranging from \SI{61}{\nano\meter} to \SI{551}{\nano\meter}. The resulting processed EEL spectra are summarised in Figure~\ref{Fig3}a. Figure~\ref{Fig3}b shows the STEM HAADF micrographs of the nanoparticles with rectangles marking the integration areas from which the EEL spectra were collected. The spectra revealed pronounced plasmon peaks corresponding to the dipole mode present in all studied nanoparticles, ranging from \SI{0.6}{\electronvolt} to \SI{3.1}{\electronvolt}. The spectra are further fitted by Gaussians to obtain the characteristic parameters of the plasmon peaks, namely, the peak energy $E$, the loss probability maxima $I_{\mathrm{max}}$, and the full-width half-maximum $\Delta E$. The results are summarised in Table~\ref{Tab1}.

\begin{table}[t]
    \centering
    \begin{tabular}{c|c|c|c|c}
         size (nm) & $E$ (eV) & $I_{\mathrm{max}}$ & $\Delta E$ (eV) & Q factor \\
         \hline
61	&	3.10	&	$1.9 \times 10^{-5}$	&	0.95	&	3.3	\\
102	&	2.66	&	$2.6 \times 10^{-5}$	&	0.80	&	3.3	\\
146	&	2.44	&	$2.9 \times 10^{-5}$	&	0.79	&	3.1	\\
205	&	1.86	&	$3.2 \times 10^{-5}$	&	0.81	&	2.3	\\
268	&	1.35	&	$3.5 \times 10^{-5}$	&	0.61	&	2.2	\\
303	&	1.16	&	$2.8 \times 10^{-5}$	&	0.47	&	2.5	\\
337	&	0.96	&	$3.5 \times 10^{-5}$	&	0.30	&	3.2	\\
410	&	0.78	&	$4.6 \times 10^{-5}$	&	0.41	&	1.9	\\
440	&	0.65	&	$4.6 \times 10^{-5}$	&	0.22	&	3.0	\\
506	&	0.58	&	$5.6 \times 10^{-5}$	&	0.22	&	2.7	\\
551	&	0.60	&	$5.0 \times 10^{-5}$	&	0.24	&	2.5	
    \end{tabular}
    \caption{Plasmonic properties of bismuth nanoparticles.}
    \label{Tab1}
\end{table}

\begin{figure}[t]
    \centering
    \includegraphics[width=1\linewidth]{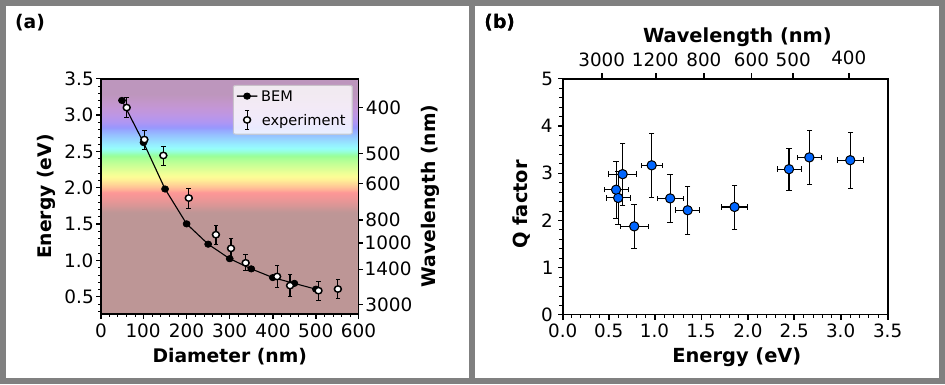}
    \caption{Spectral tunability of bismuth nanoparticles: (a) Dipole LSPR energy extracted from the measured and calculated EEL spectra as a function of nanoparticle diameter. (b) The Q-factors of the dipole plasmon mode extracted from the fits of the LSPR peaks in the measured EEL spectra.}
    \label{Fig4}
\end{figure}

The spectral tunability of the dipole plasmon mode is shown in Figure~\ref{Fig4}a. Within the range of investigated nanoparticle diameters, the dipole mode energy spans from the near-infrared to the ultraviolet part of the spectrum. The energy of the dipole mode covers the interval from \SI{0.60}{\electronvolt} (corresponding to the wavelength of \SI{2066}{\nano\meter}) for the \SI{551}{\nano\meter} nanoparticle, to \SI{3.10}{\electronvolt} (\SI{400}{\nano\meter} in wavelength) for the \SI{61}{\nano\meter} nanoparticle. Therefore, bismuth nanoparticles represent a biocompatible and oxidation-resistant hyperspectral plasmonic platform that is tunable from the near-infrared to the ultraviolet part of the spectrum. This property is analogous to that of gallium \cite{10.1021/acs.jpclett.3c00094} or silver amalgam \cite{10.1021/acs.jpcc.9b04124} nanoparticles.

Additionally, we have performed numerical simulations (see Methods) to support our experimental results. The calculated EEL spectra are shown in Figure~S5. The energy of the dipole mode, extracted from the simulations, is also shown in Figure~\ref{Fig4}a. We generally see a good agreement between the experiment and the numerical simulations. Small differences suggest a minor inaccuracy in the parameters used in the numerical model, including the exact shape of the nanoparticles, the dielectric function of bismuth, or the approximation of a silicon dioxide membrane by an effective surrounding medium.

The highest loss probabilities of the dipole mode were observed in nanoparticles with diameters ranging from \SI{410}{\nano\meter} to \SI{551}{\nano\meter}, where the loss probability fluctuated around $5 \times 10^{-5}$. For smaller nanoparticles, the loss probability decreases with the decreasing diameter of the nanoparticle, reaching the lowest loss probability of $1.9 \times 10^{-5}$ for the \SI{61}{\nano\meter} nanoparticle. The FWHM of the dipole LSPR mode exhibits the opposite trend. It increases with decreasing nanoparticle diameter and reaches \SI{0.95}{\electronvolt} for the smallest (\SI{61}{\nano\meter}) nanoparticle and \SI{0.22}{\electronvolt} for the largest (\SI{551}{\nano\meter}) nanoparticle, respectively. The resulting Q factors range from 1.9 (for the \SI{410}{\nano\meter} nanoparticle) to 3.3 (\SI{102}{\nano\meter} nanoparticle). They are shown in Figure~\ref{Fig4}b as a function of the energy of the dipole mode. When experimental uncertainty is taken into account, the Q factors are comparable for all nanoparticles analysed, promising a comparable plasmonic performance over the entire available spectral range.

In conclusion, we have synthesised spherical monocrystalline bismuth nanoparticles from sodium bismuthate in ethyleneglycol through a solvothermal surfactant-free process. The nanoparticles were characterised using STEM-EELS to study their plasmonic properties at the single-particle level and to investigate the spectral tunability of localised surface plasmon resonances as a function of the nanoparticle diameter. The absence of surfactant molecules helped to avoid the influence of the EELS study by the locally different dielectric environment, enabling an unaffected study of the plasmonic response of individual nanoparticles. The results demonstrate that the absence of surfactants does not adversely affect the structure of the synthesised nanoparticles. The tunability of dipole plasmon resonant modes is demonstrated through the nanoparticle diameter, covering the interval from the near-infrared to the ultraviolet spectral region. Furthermore, plasmonic resonances demonstrate stability across the entire spectral bandwidth, thereby enhancing the attractiveness of bismuth nanoparticles for applications over a wide spectral range and establishing them as a viable option for plasmonics. Moreover, the lower cost of bismuth, in conjunction with its biocompatibility and resistance to oxidation, renders it a suitable candidate for utilisation, particularly in industrial and large-scale plasmonic applications.

\section{Methods}

\subsubsection{Sythesis of bismuth nanoparticles}

Monocrystalline bismuth nanoparticles were synthesised using a modified polyol process based on the synthesis approach reported in Ref. \cite{10.1021/jp063474e}. We dissolved \SI{0.30}{\gram} of \ce{NaBiO3} in \SI{30}{\milli\litre} of ethyleneglycol. We then placed the solution in a sand bath and gradually heated it to \SI{197}{\celsius} while stirring at a rate of 400\,RPM. When the temperature reached approximately \SI{170}{\celsius}, the bright yellow solution turned dark orange. After maintaining the temperature of \SI{197}{\celsius} for around 20 minutes, we switched off the heating and left the solution to slowly cool to room temperature.

\subsubsection{Analytical transmission electron microscopy}

TEM measurements, including STEM HAADF imaging and atomic resolution TEM, were performed on a TEM FEI Titan operated at \SI{300}{\kilo\electronvolt}. EELS measurements were performed on a TEM FEI Titan equipped with a GIF Quantum spectrometer operated at \SI{120}{\kilo\electronvolt} in the scanning monochromated mode with the convergence semi-angle set to 10\,mrad and the collection semi-angle set to 11.4\,mrad. The probe current was adjusted to around \SI{100}{\pico\ampere}. The dispersion of the spectrometer was set to \SI{0.01}{\electronvolt} per channel and the FWHM of the zero-loss peak was around \SI{0.15}{\electronvolt}. The acquisition time was adjusted to use the maximal intensity range of the CCD camera in the spectrometer and avoid its overexposure. These parameters were selected to acquire the EELS signal with the highest signal-to-background ratio \cite{10.1016/j.ultramic.2020.113044}. EEL spectra were integrated over rectangular areas at the edges of the nanostructures where the LSPR is significant. They were further divided by the integral intensity of the zero-loss peak to transform the measured counts into a quantity proportional to the loss probability, background subtracted by subtracting the EEL spectrum of a pure silicon nitride membrane, and fitted by Gaussians.

\subsubsection{Numerical simulations}

EEL spectra were calculated using the MNPBEM toolbox \cite{10.1016/j.cpc.2015.03.023} based on the boundary element method. Our model consisted of a bismuth sphere in an effective surrounding medium. The dielectric function of bismuth was taken from Ref. \cite{10.1063/1.3243762} and the effective refractive index of the surrounding medium was set to 1.3 to approximate the effect of the silicon dioxide membrane substrate. The \SI{120}{\kilo\electronvolt} electron beam was positioned \SI{30}{\nano\meter} outside the nanoparticle.

\begin{acknowledgement}

This work is supported by the project QM4ST (project No. CZ.02.01.01/00/22\_008/0004572) by OP JAK, call Excellent Research, project Czech-NanoLab by MEYS CR (project No. LM2023051), and Brno University of Technology (project No. FSI-S-23-8336). M.F. acknowledges the support of the Brno Ph.D. talent scholarship.

\end{acknowledgement}

\begin{suppinfo}

Supporting Information: STEM HAADF micrographs of bismuth nanoparticles capturing the occasional nonsperical shapes, the influence of the synthesis time on the resulting size of bismuth nanoparticles, contamination of the bismuth nanoparticles by residues imaged by STEM and analysed by STEM EDX, and  calculated EEL spectra of bismuth nanospheres (PDF).

\end{suppinfo}

\bibliography{reference}

\end{document}